\documentclass[twocolumn,superscriptaddress]{revtex4}
\def\bibcommenthead{}

\usepackage{times}
\usepackage{braket}
\usepackage{xcolor}
\usepackage{graphicx}
\usepackage{multirow}
\usepackage{hyperref}

\begin{document}

\title{Floquet-Tailored Rydberg Interactions} 
	\author{Luheng Zhao}
	\affiliation{Centre for Quantum Technologies, National University of Singapore, 117543 Singapore, Singapore}
\author{Michael Dao Kang Lee}
	\affiliation{Centre for Quantum Technologies, National University of Singapore, 117543 Singapore, Singapore}
\author{Mohammad Mujahid Aliyu}
	\affiliation{Centre for Quantum Technologies, National University of Singapore, 117543 Singapore, Singapore}
\author{Huanqian Loh}
        \email[]{phylohh@nus.edu.sg}
	\affiliation{Centre for Quantum Technologies, National University of Singapore, 117543 Singapore, Singapore}
        \affiliation{Department of Physics, National University of Singapore, 117542 Singapore, Singapore}

% \date{}

%%%%%%%%%%%%%%%%% ABSTRACT %%%%%%%%%%%%%%%%

\begin{abstract}
The Rydberg blockade is a key ingredient for entangling atoms in arrays. However, it requires atoms to be spaced well within the blockade radius, which limits the range of local quantum gates. Here we break this constraint using Floquet frequency modulation, with which we demonstrate Rydberg-blockade entanglement beyond the traditional blockade radius and show how the enlarged entanglement range improves qubit connectivity in a neutral atom array. Further, we find that the coherence of entangled states can be extended under Floquet frequency modulation. Finally, we realize Rydberg anti-blockade states for two sodium Rydberg atoms within the blockade radius. Such Rydberg anti-blockade states for atoms at close range enables the robust preparation of strongly-interacting, long-lived Rydberg states, yet their steady-state population cannot be achieved with only the conventional static drive. Our work transforms between the paradigmatic regimes of Rydberg blockade versus anti-blockade and paves the way for realizing more connected, coherent, and tunable neutral atom quantum processors with a single approach.
\end{abstract}

\maketitle

%%%%%%%%%%%%%%%%%%%% INTRODUCTION %%%%%%%%%%%%%%%%%%%%
%\section*{Introduction}
Ultracold atoms in reconfigurable tweezer arrays have emerged as one of the most powerful and rapidly growing quantum platforms. These systems have demonstrated impressive quantum many-body simulations \cite{ebadi2021quantum, scholl2021quantum, yan2022two}, highly stable frequency standards \cite{young2020half, madjarov2019atomic}, and promising quantum computation architectures \cite{kaufman2015entangling, saffman2016quantum, jau2016entangling, levine2019parallel, madjarov2020high, bluvstein2022quantum, chew2022ultrafast, mcdonnell2022demonstration, ma2022universal, jenkins2022ytterbium, okuno2022takahashi, singh2022dual, saw2020ultracold, zhang2022optical}.

At the heart of these quantum applications lies entanglement, which is often effected in neutral atom arrays via the Rydberg blockade \cite{urban2009observation, gaetan2009observation}. Among the various entanglement schemes \cite{jaksch2000fast, jau2016entangling, levine2019parallel, madjarov2020high, jo2020rydberg}, the Rydberg blockade has been widely adopted due to its robustness against position disorder. However, it requires the distance between two atoms to be well within the blockade radius to prevent substantial entanglement infidelity due to blockade error. This constraint reduces quantum gates on Rydberg atoms to a limited range. Improvements in the Rydberg interaction range would increase the qubit connectivity, which could significantly enhance quantum processing efficiency.

Like other quantum processors, neutral atom computations and simulations are limited by decoherence \cite{de2018analysis}. New methods for extending the lifetime of entangled states would improve the fidelity of quantum gates \cite{saffman2016quantum}. However, even with existing levels of decoherence, neutral atom processors continue to demonstrate excellent quantum simulations \cite{ebadi2021quantum, scholl2021quantum, yan2022two}. As these systems continue to develop more robust and efficient ways to initialize quantum states, they are likely to yield more diverse quantum simulation capabilities.

In this work, we report that these neutral atom platforms can be advanced on three critical fronts --- extending the blockade-based entanglement range, improving coherence times, and enabling 
versatile state-preparation schemes --- with Floquet frequency modulation (FFM). Our FFM approach is simple and straightforward to implement in all existing neutral atom array experiments. First, we demonstrate that atoms can be entangled outside the traditional blockade radius, as predicted by Refs \cite{basak2018periodically, mallavarapu2021population} , thereby significantly increasing the useful range of the Rydberg interaction. Furthermore, we show how FFM can protect a two-atom entangled state against Doppler dephasing, which is the typical mechanism limiting entangled-state coherence in a Rydberg atom array. Finally, we propose a robust transfer of closely-spaced atoms from the ground state into an anti-blockaded state. Such a strongly-interacting state cannot be otherwise attained in the steady state with a conventional static drive, yet its realization would open the door to intricate simulations of quantum dynamics.

A versatile approach, FFM has been used to implement efficient random access quantum processors with superconducting circuits \cite{naik2017random}, generate strongly interacting polaritons with Rydberg atoms in an optical cavity \cite{clark2019interacting}, and create exotic states of light \cite{clark2020observation}. FFM is also analogous to the shaking of optical lattices, which has been used to realize synthetic gauge fields in ultracold atoms \cite{aidelsburger2013realization, miyake2013realizing, eckardt2017colloquium}. In a Rydberg atom array, FFM has been used to stabilize the revivals of quantum many-body scars \cite{bluvstein2021controlling}. 

In this work, we focus on the entanglement between two atoms (Fig.~1), which serves as a basic ingredient of quantum information processing. In addition, the results obtained here can be generalized to a large atom array by using single-site addressing lasers to select two atoms at a given time.

%%%%%%%%%%%%%%%%%%%% FFM - MODEL AND IMPLEMENTATION %%%%%%%%%%%%%%%%%%%%
%\section*{Results}

\paragraph*{Floquet frequency modulation: model and implementation.}
When atoms are excited from the ground state $\ket{g}$  to the Rydberg state $\ket{e}$, the resulting dynamics are governed by the Hamiltonian:
\begin{equation} \label{eq:Hnative}
\frac{H}{\hbar} = -\Delta(t) \sum_{i=1}^{N} \sigma_{ee}^{i} + \frac{\Omega}{2}\sum_{i=1}^{N} \sigma_{x}^{i} + \sum_{i<j} V_{ij} \sigma_{ee}^{i} \sigma_{ee}^{j} \, ,
\end{equation}
where $i$ indexes the atom, $V(r) = C_6/r^6$ is the van der Waals interaction between Rydberg atoms, and $\Omega$ is the Rabi frequency. Under FFM, the laser detuning $\Delta(t)$ is modulated sinusoidally in time with modulation amplitude $\delta$ and modulation frequency $\omega_0$ about an offset $\Delta_0$ to give $\Delta(t)= \Delta_0 + \delta \sin (\omega_0 t)$.

\begin{figure}[t] 	
    \centering
	\includegraphics[keepaspectratio,width=8.8cm]{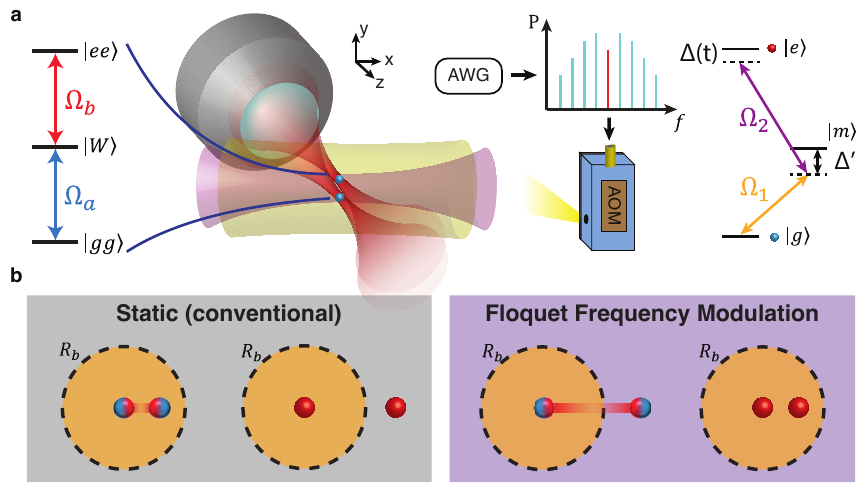}
        \caption{\textbf{Using Floquet frequency modulation to transform Rydberg interactions.} \textbf{a} Experiment setup for implementing FFM on two tweezer-trapped $^{23}$Na atoms with counter-propagating Rydberg lasers at 589 nm ($\Omega_{1}$) and 409 nm ($\Omega_{2}$). The 589~nm Rydberg laser is frequency-modulated with an acousto-optic modulator (AOM) driven by an arbitrary waveform generator (AWG). \textbf{b} Transforming between the two regimes of interactions: Rydberg blockade (two entangled atoms depicted as blue/red spheres) versus anti-blockade (two red atoms). With the conventional static drive, the Rydberg-blockade regime can only be accessed with the atoms positioned well within the blockade radius $R_{b}$, while atoms spaced farther than the blockade radius experience anti-blockade. The counter-intuitive regime of interactions can be exploited with Floquet frequency modulation, where atoms outside the static blockade radius experience Rydberg blockade and vice versa, as predicted by Refs \cite{basak2018periodically}, \cite{mallavarapu2021population}.} 
    \label{fig:ffm}
\end{figure}

For the resonant addressing of an atom-array building block comprising two atoms, we set $\Delta_0 = 0$ and $N = 2$. In the absence of dissipation, the two-atom system can evolve between the $\ket{gg}$, $\ket{W} = (\ket{ge} + e^{i\phi}\ket{eg})/\sqrt{2}$, and $\ket{ee}$ states, where $\phi$ denotes the relative phase between the atoms arising from their initial positions. With the application of FFM, the two-atom Hamiltonian can be transformed to a new Hamiltonian in a rotating frame, such that the coupling strengths for the $\ket{gg} \leftrightarrow \ket{W}$ and $\ket{W} \leftrightarrow \ket{ee}$ transitions are respectively rescaled to \cite{basak2018periodically, mallavarapu2021population}:
\begin{eqnarray} 
\label{eq:omegaa} 
\Omega_{a}(t) &\propto& \Omega\sum_{m=-\infty}^{\infty} J_{m}(\alpha) e^{i m \omega_0t + im\frac{\pi}{2}} \, , \\
\label{eq:omegab}
\Omega_{b}(t) &\propto& \Omega\sum_{m=-\infty}^{\infty} J_{m}(\alpha) e^{i(m\omega_{0} + V)t + im\frac{\pi}{2}} \, ,
\end{eqnarray}
where $J_{m}(\alpha)$ is the $m^{th}$ order Bessel function of the first kind with modulation index $\alpha = \delta/\omega_{0}$.

A high modulation frequency simplifies the picture as only the resonant terms dominate the coupling strength. Explicitly, the Rabi frequency for $\ket{gg} \leftrightarrow \ket{W}$ is dominated by $J_0(\alpha)$, whereas the resonance condition for $\ket{W} \leftrightarrow \ket{ee}$ is met by setting $m\omega_{0} = -V$. The interplay of these two resonance conditions dictates the physics behind the results presented in this study.

To implement FFM, we send a Rydberg excitation laser through an acousto-optic modulator driven with a time-varying frequency (Fig.~1a). Since the dynamics of the two-atom system depends critically on the modulation index $\alpha$, we take care to calibrate the modulation amplitude $\delta$, which can differ from the specified amplitude due to the finite modulator bandwidth. We further minimize the residual amplitude modulation arising from the frequency-dependent diffraction efficiency of the acousto-optic modulator (see Supplementary Information). We note that FFM can be easily implemented given the typical bandwidth of commercially available acousto-optic modulators. 

We point out that FFM is distinct from the Floquet engineering techniques used to control dynamics in Rydberg atoms \cite{borish2020transverse, geier2021floquet, scholl2022microwave}. The latter involves changing the effective Hamiltonian by applying fast, periodic pulses with well-defined phase relations. These pulses are used in the microwave domain to address either two ground states in a Rydberg-dressed system or two Rydberg states. In contrast, FFM directly transforms the effective couplings. Further, FFM does not use periodic pulses, which would have been challenging to implement in the optical domain on the time scale of Rabi oscillations between the ground and Rydberg states.

Our experiment procedure begins with the $D_1$ $\Lambda$-enhanced loading of single $^{23}$Na atoms into two optical tweezers \cite{aliyu2021d, brown2019gray, angonga2022gray}. We excite the $^{23}$Na atoms from the ground state $\ket{g} =$ $|3S_{1/2}, F=2,$ $\left.m_{F} = 2\right\rangle$ to the Rydberg state $\ket{e} = \ket{59S_{1/2}, m_{J} = 1/2}$ via the intermediate state $\ket{m} = \ket{3 P_{3/2}, F = 3, m_{F} = 3}$ with two photons at 589~nm and 409~nm. The Rydberg laser intensities and single-photon detuning $\Delta'$ are chosen to give an effective single-atom Rabi frequency of $\Omega/(2\pi) = 1.0$~MHz. FFM is only applied to the 589~nm excitation laser. The optical tweezers are switched off during Rydberg excitation and turned back on at the end of the excitation sequence to image the atoms. Using $D_1$ imaging, the ground (Rydberg) state is detected as the presence (absence) of a recaptured atom. The false positive error (3\%) is dominated by the imaging survival probability, whereas the false negative error due to the decay of Rydberg atoms is estimated to be 3\%.

%%%%%%%%%%%%%%%%%%%%%%%% Extended Rydberg blockade range %%%%%%%%%%%%%%%%%%%%%%%%
\paragraph*{Extended Rydberg blockade range.} For a static Rydberg excitation $(\alpha = 0)$, suppression of population in the $\ket{ee}$ state is an important prerequisite for the high-fidelity generation of the entangled $\ket{W}$ state. Therefore, quantum gates are typically carried out between atoms spaced well within the Rydberg blockade radius $R_b$, where $V(R_{b}) = \Omega$. Under FFM, the two-atom couplings are rescaled by Bessel functions (Eqs.~(\ref{eq:omegaa}) and (\ref{eq:omegab})), giving an intuitive picture for the transformation of the Rydberg blockade radius. 

\begin{figure*}[!htbp]
	\includegraphics[keepaspectratio,width=18cm]{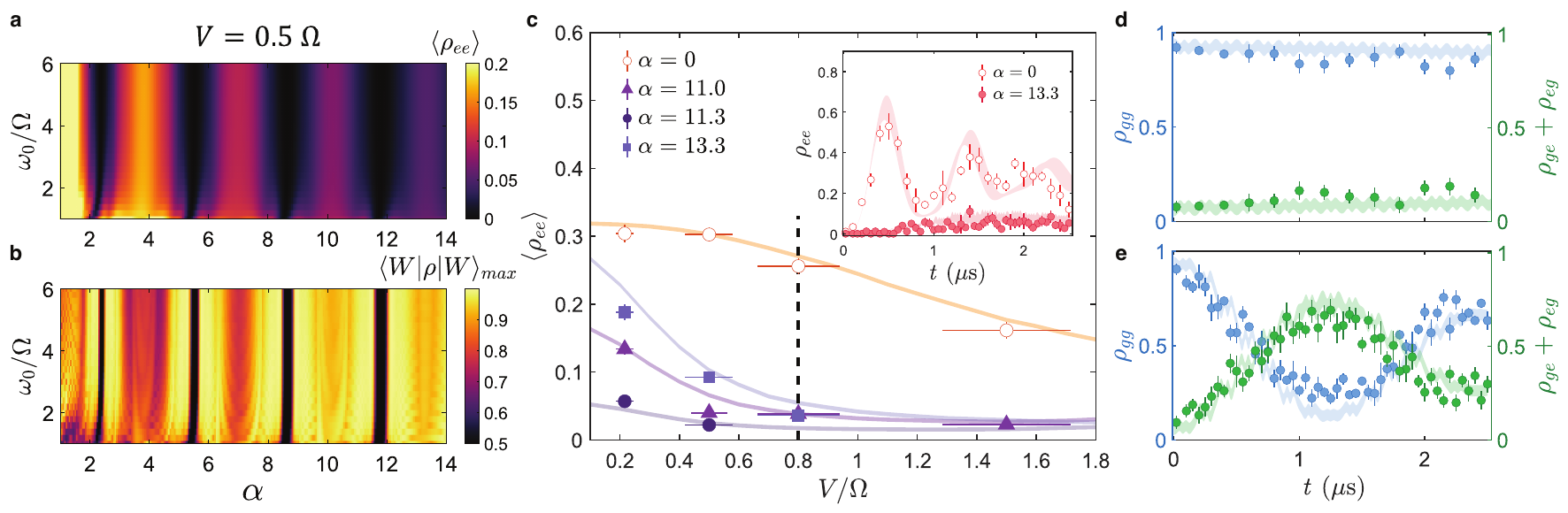}
        \caption{\textbf{Extended Rydberg blockade range.} \textbf{a} Calculated $\ket{ee}$ population, time averaged over 5~$\mu$s, and \textbf{b} calculated maximum $\ket{W}$ fidelities as a function of normalized modulation frequency $\omega_{0}/\Omega$ and modulation index $\alpha$. \textbf{c} Measured $\ket{ee}$ population, time-averaged over 2.5~$\mu$s, as a function of normalized interaction strength $V/\Omega$ for $\omega_{0} = 3~\Omega$ and different modulation indices $\{\alpha\}$. Horizontal error bars are attributed to the position uncertainty of the atoms. Solid lines indicate the numerical modeling results for the corresponding $\alpha$. (Inset) Population dynamics $\rho_{ee}$ at $V= 0.8~\Omega$. \textbf{d, e} Population dynamics for (left axis, blue) $\rho_{gg}$ and (right axis, green) $\rho_{eg} + \rho_{ge}$, measured at $V = 0.8~\Omega$ and $\omega_{0} = 3~\Omega$. The shaded curves reflect the results of Monte Carlo simulations. \textbf{d} Population trapping in $\ket{gg}$ is observed for $\alpha = 5.5$, where $J_{0}(\alpha) \approx 0$. \textbf{e} At $\alpha = 6.9$, the $\ket{W}$ state is generated with a calculated maximum fidelity of 0.77(5). In panels (\textbf{c})--(\textbf{e}), the displayed vertical error bar associated with each data point reflects the $1\sigma$ confidence interval.} 
	\label{fig:blockade}
\end{figure*}

As an example, we consider two atoms spaced farther than the static blockade radius, such that their interaction strength is given by $V(r) = 0.5~\Omega$. Figures~2a and 2b show the calculated time-averaged $\ket{ee}$ populations and the maximum $\ket{W}$ fidelities, respectively, as a function of the modulation frequency and modulation index. In the regime of high modulation frequency $(\omega_0 \geq 2~\Omega)$, the dynamics are relatively robust and are dictated by the Bessel functions. At the Bessel function zeros ($J_{0}(\alpha) = 0$), the $\ket{ee}$ population is the most suppressed. However, it would be misleading to use these Bessel function zeros for generating entangled states, as the population is actually trapped in its initial $\ket{gg}$ state. To generate entangled states, it is optimal to select $\alpha$ slightly away from the Bessel function zeros. Here, the rescaled Rabi frequency $\Omega_a \propto J_0(\alpha)\Omega$ remains finite but small compared to the interaction, such that errors arising from populating the $\ket{ee}$ state can still be suppressed while the $\ket{W}$ state can be generated with good fidelity. Furthermore, by operating near a higher order Bessel function zero, the $\ket{W}$ state can be realized with high fidelity over a wider range of modulation indices. Given the finite modulation bandwidth of the acousto-optic modulator, it is advantageous to work with a modulation frequency that is large enough compared to the Rabi frequency, but still low enough to allow a high modulation index (e.g.~$\omega_{0} = 3~\Omega$). 

Figure~2c shows the measured blockade enhancement under FFM, taken with $\omega_{0} = 3~\Omega$ and at various modulation indices. Compared to the static excitation, the dynamics of the $\ket{ee}$ state under FFM are significantly suppressed (inset, $V = 0.8~\Omega$). The observed Rydberg blockade over a range of atom spacings ($r/R_{b} = 0.93 - 1.3$, inferred from the normalized interaction range $V/\Omega  =0.22 - 1.5$) agrees well with the theory simulations. Through an appropriate choice of $\alpha$, we observe either population trapping (Fig.~2d, $\alpha = 5.5$) or coherent dynamics between the $\ket{gg}$ and $\ket{W}$ states (Fig.~2e, $\alpha = 6.9$). The $\ket{W}$  fidelity achieved in Fig.~2e is determined from Monte Carlo simulations to be 0.77(5). The observed fidelity is primarily limited by the coherence of the Rydberg excitation lasers, which can be improved with cavity filtering techniques \cite{levine2018high}.

Where the coherence of the Rydberg excitation laser and the finite lifetime of the Rydberg state are no longer dominant constraints, the entangled state fidelity can be optimized by choosing $\alpha$ in exchange for longer gate times. For instance, one can use FFM to access a $\ket{W}$ fidelity of $0.98$ using $\alpha = 11.1$ even at a small interaction strength of $V = 0.5~\Omega$. To access the same $\ket{W}$ fidelity with the static Rydberg excitation scheme, the atoms would have had to experience an interaction strength of $V = 4.9~\Omega$, effectively extending the Rydberg blockade range by a factor of $(4.9/0.5)^{1/6} > \sqrt{2}$ (see Supplementary Information). In other words, for an atom array with a fixed square geometry, the static scheme would have allowed only the four nearest neighbors to be entangled in a pairwise manner with the center atom, whereas implementing FFM would allow the next-nearest neighbors on the diagonals to be pairwise entangled with the center atom, thereby potentially doubling the number of qubit connections on demand.

In the rest of this paper, we switch to working well within the Rydberg blockade radius and demonstrate two more useful features of the FFM.

%%%%%%%%%%%%%%%%%%%%%%%% Protection of entangled state lifetime %%%%%%%%%%%%%%%%%%%%%%%%
\paragraph*{Protection of entangled state against dephasing.} The $\ket{W}$ state can be dynamically stabilized \cite{mallavarapu2021population} under the same conditions that give rise to population trapping $(J_{0}(\alpha) = 0)$. We demonstrate this by first transferring atoms to the $\ket{W}$ state with a static, resonant $\pi$-pulse in the Rydberg blockade regime ($V = 8~\Omega$). Subsequently, we apply the FFM ($\omega_{0} = 6~\Omega, \alpha = 5.5$) for 2~$\mu$s, before returning to the static drive. The Rabi frequency is kept constant throughout the sequence (Fig.~3a). During the FFM, the dynamics of the $\ket{W}$ state are frozen (Fig.~3b), in contrast to the case where the static drive is applied throughout the sequence (Fig.~3c). 

\begin{figure*}[!htbp]
	\centering
	\includegraphics[width=18cm]{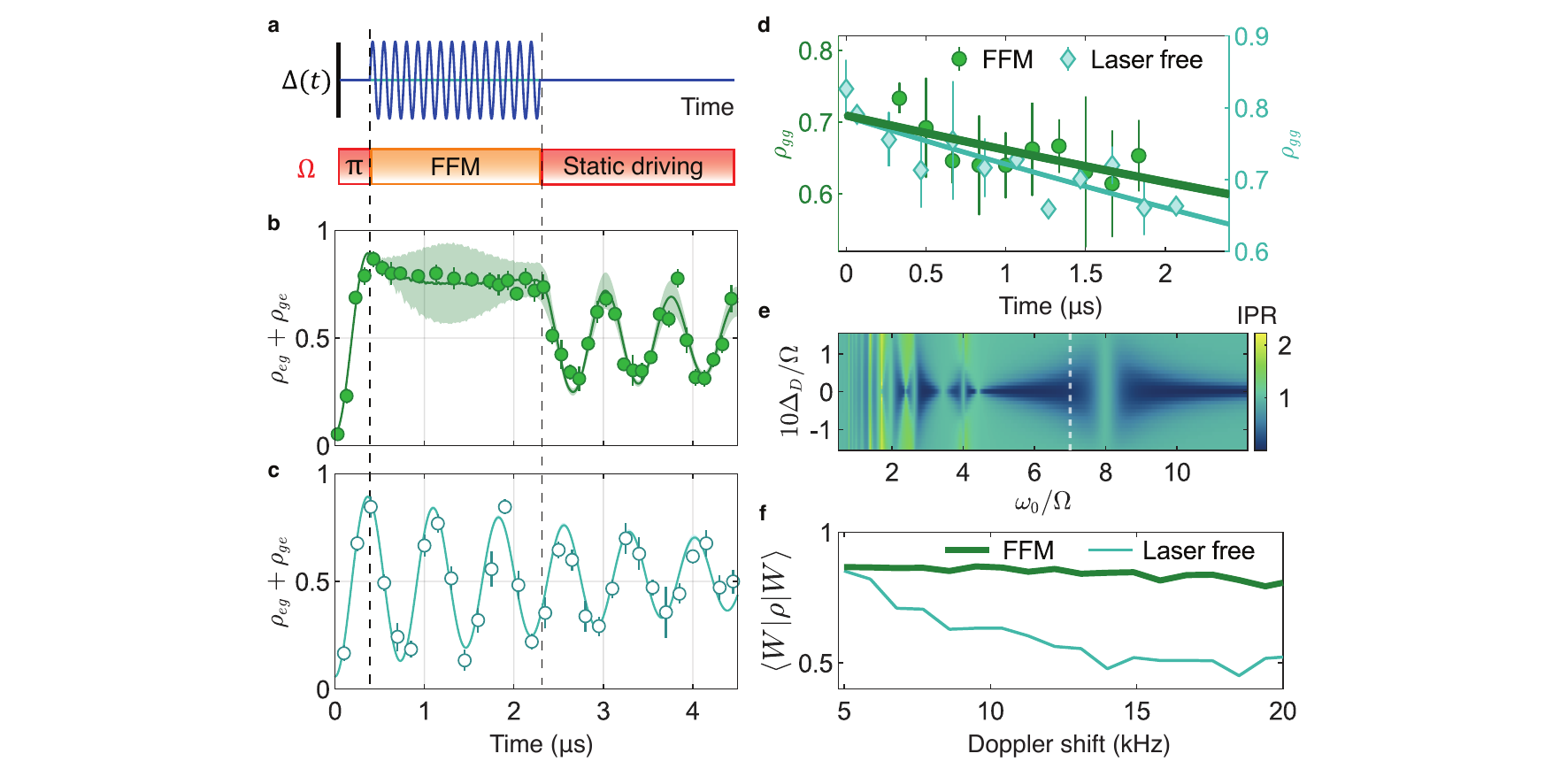}
        \caption{\textbf{Dynamical stabilization of entangled states at the Bessel function zero $\alpha = 5.5$ and at $V = 8~\Omega$.} \textbf{a} Pulse sequence comprising an initial $\pi$ pulse, $2~\mu$s of FFM ($\omega_{0} = 6~\Omega$), followed by static driving. \textbf{b, c} Observed population dynamics $\rho_{eg} + \rho_{ge}$ \textbf{b} under the above FFM pulse sequence versus \textbf{c} under a continuous, resonant static drive. The solid lines account for the decoherence effects as modeled by Lindblad superoperators. The shaded bands depict Monte Carlo simulations that include the atom position uncertainty. \textbf{d} Comparison of $\ket{W}$ decay for (left axis, green) FFM versus (right axis, blue) a laser-free evolution. The FFM evolution yields a decay time of 14(5)~$\mu$s whereas the laser-free evolution yields a decay time of 11(2)~$\mu$s. Error bars in panels (\textbf{b})--(\textbf{d}) depict the $1\sigma$ confidence interval. \textbf{e} Calculated IPR as a function of rescaled Doppler shift $10\Delta_{D}/\Omega$ and normalized modulation frequency $\omega_{0}/\Omega$. (White dashed line) At $\omega_{0} = 7~\Omega$, the IPR is approximately 0 over an extended range of Doppler shifts, which is desired for robust dynamical stabilization of $\ket{W}$. \textbf{f} Calculated $\ket{W}$ fidelity for different Doppler shift widths after 20~$\mu$s of (green) FFM or (blue) laser-free evolution. Here the FFM parameters are $\omega_{0} = 7~\Omega, \alpha = 5.5$.}
	\label{fig:entprotection}
\end{figure*}

Instead of FFM, one can also trivially keep atoms in the $\ket{W}$ state by turning off the excitation lasers after the first $\pi$-pulse. We refer to this alternative as the laser-free scheme. In each case (laser-free versus FFM), the entangled state coherence is limited by relative Doppler shifts between the two atoms \cite{levine2018high}, and can be measured by first applying a $\pi$-pulse to drive $\ket{gg}$ to $\ket{W}$, then waiting for a variable time $0 \leq t \leq 2$~$\mu$s before applying another $\pi$-pulse to measure the population in $\ket{gg}$. Figure~3d compares the measured decay of the $\ket{W}$ state for both cases, where the FFM sequence yields a fit decay time of 14(5)~$\mu$s and the laser-free scheme shows a comparable decay time of 11(2)~$\mu$s. 

With a judicious choice of parameters, FFM can protect the entangled $\ket{W}$ state from dephasing and maintain its coherence over the laser-free case. Intuitively, the protection arises from a high-frequency interruption of the dephasing process. More specifically, although we null out the Rabi coupling between $\ket{gg}$ and $\ket{W}$ with FFM, the off-resonant coupling between $\ket{W}$ and $\ket{ee}$ remains. Since the $\ket{ee}$ state is robust against drifts in relative phase shifts between the atoms, the Doppler dephasing is periodically interrupted whenever the system evolves to pick up the small $\ket{ee}$ admixture. The fast residual oscillation is washed out in a time average, giving rise to a net protection of the $\ket{W}$ state.

To access the entanglement-protection regime, one needs to ensure that the $\ket{W}$ state has a strong overlap with an eigenstate of the Floquet Hamiltonian $\ket{\phi_{k}(0)}$⟩. The overlap $p_{k} = |\braket{\phi_{k}(0)|W}|^{2}$ is parameterized by the inverse participation ratio (IPR), which is defined \cite{mallavarapu2021population} to be $\Pi^{\ket{W}} = (1/\sum_{k} p_{k}^{2}) - 1$. A low IPR is preferred for robust dynamical stabilization of the $\ket{W}$ state. Operating, for instance, at $\omega_{0} = 7~\Omega$ yields a low IPR over an extended range of Doppler shifts (Fig.~3e). After 20~$\mu$s of FFM application, the $\ket{W}$ state fidelity under FFM is predicted to be significantly higher than that for the laser-free case (Fig.~3f). The higher fidelity under FFM is maintained over a large range of Doppler shifts, demonstrating the robustness of entanglement under FFM.

%%%%%%%%%%%%%%%%%%%%%%%% ANTI-BLOCKADE %%%%%%%%%%%%%%%%%%%%%%%%%%
\paragraph*{Enhanced Rydberg anti-blockade dynamics.} We now turn our attention to Rydberg anti-blockade states \cite{amthor2010evidence}, which are promising for quantum simulations of interesting dynamics such as that of epidemics \cite{wintermantel2021epidemic} and of flat band systems in condensed matter physics \cite{liu2022localization}. Rydberg anti-blockade (i.e.~$\ket{ee}$ population) is typically achieved for two atoms spaced outside the blockade radius \cite{jo2020rydberg}. Within the blockade radius, Rabi oscillations between the $\ket{gg}$ and $\ket{ee}$ states can still be realized \cite{marcuzzi2017facilitation} with a finite static detuning, e.g.\ $\Delta_{0} = V/2$. However, this comes at the expense of slower dynamics, particularly when $V \gg \Omega$, as the effective Rabi frequency is set by $\Omega^{2}/\Delta_{0}$.

\begin{figure*}[!htbp]
	\centering
	\includegraphics[keepaspectratio,width=18cm]{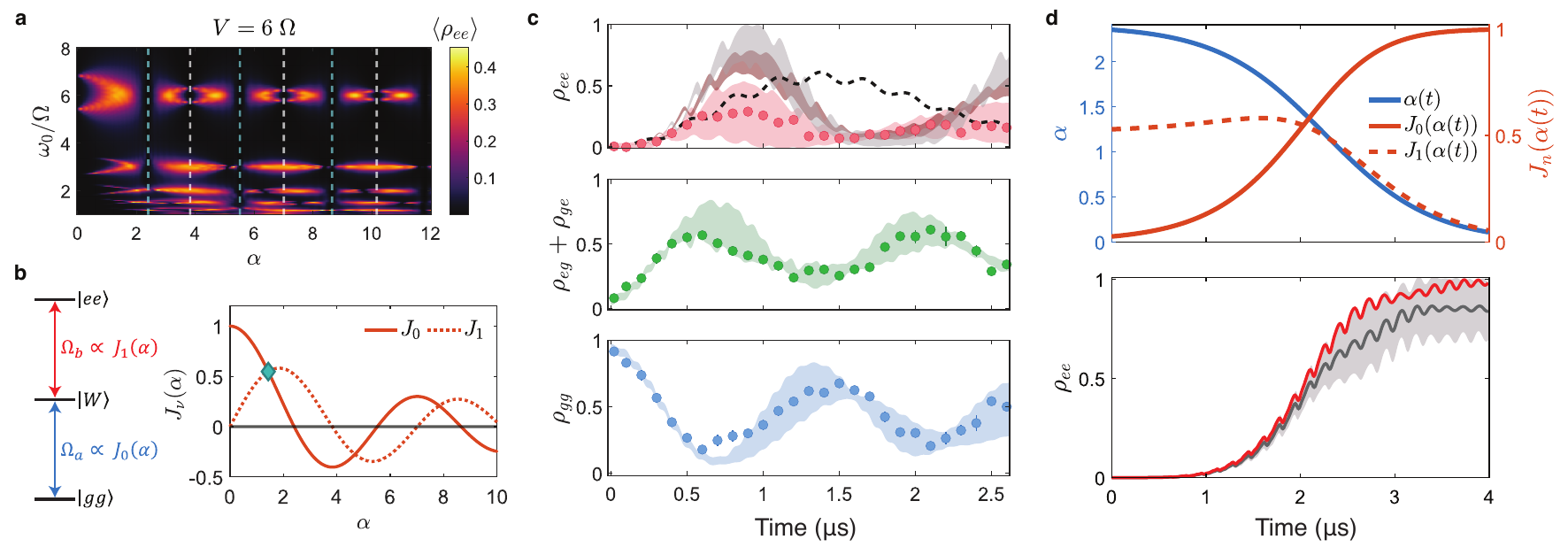}
        \caption{\textbf{Enhanced Rydberg anti-blockade dynamics at $V = \omega_{0} = 6~\Omega$.} \textbf{a} Calculated $\ket{ee}$ population, time-averaged over $10~\mu$s, as a function of normalized modulation frequency $\omega_{0}/\Omega$ and modulation index $\alpha$. The green and white dashed lines indicate positions at which $J_{0}(\alpha) = 0$ and $J_{1}(\alpha) = 0$, respectively. \textbf{b} The two-atom Rabi frequencies $\{\Omega_{a}, \Omega_{b}\}$ depend strongly on the modulation index. (Green diamond) At $\alpha = 1.4$, both Rabi frequencies have large values and can be used to access Rydberg anti-blockade states. \textbf{c} Measured population dynamics (top) $\rho_{ee}$, (middle) $\rho_{eg} + \rho_{ge}$, and (bottom) $\rho_{gg}$ for $\alpha = 1.4$, which are in good agreement with Monte Carlo simulations represented by shaded curves (in same color as data points). Error bars represent the $1\sigma$ confidence interval. (Brown shading) The FFM-induced antiblockade population can be further increased with the help of ground-state cooling. The gray shading indicates the simulated $\rho_{ee}$ achieved by FFM under both ground-state cooling and enhanced coherence times of 74~$\mu$s. In general, the FFM dynamics are faster than that with an off-resonant static drive (black dashed line), where $\Delta_{0} = V/2$. \textbf{d} Simulation of steady-state Rydberg anti-blockade, achieved by combining FFM with STIRAP. (Top) The modulation index $\alpha$ is smoothly varied from 2.4 to 0, such that the two-atom Rabi frequencies change in time despite holding $\Omega$ constant. (Bottom) Calculated population dynamics $\rho_{ee}$ for the proposed STIRAP sequence. (Red line) In the absence of imperfections, $\rho_{ee}$ can be as high as 0.98. (Gray line) With ground state cooling and enhanced coherence times of 74~$\mu$s, the mean steady-state population becomes $\rho_{ee} = $ 0.85, with one standard deviation of the Monte Carlo sample distribution depicted as the half width of the light gray band.} 
	\label{fig:antiblockade}
\end{figure*}

On the other hand, FFM provides a convenient handle to access the $\ket{ee}$ state from the $\ket{gg}$ state by fulfilling both resonance conditions\cite{basak2018periodically, mallavarapu2021population} described in Eqs.~(\ref{eq:omegaa}) and (\ref{eq:omegab}) (Fig.~4a). For instance, setting $\omega_{0} = V$ causes the coupling strength for the $\ket{W} \leftrightarrow \ket{ee}$ transition to be dominated by $J_{1}(\alpha)$, while $J_{0}(\alpha)$ continues to dominate the $\ket{gg} \leftrightarrow \ket{W}$ coupling strength. Consequently, choosing a modulation index that gives a large value for both $J_1(\alpha)$ and $J_0(\alpha)$, such as $\alpha = 1.4$, realizes the Rydberg anti-blockade (Fig.~4b). Figure~4c shows the corresponding two-atom population dynamics for $V = \omega_0 = 6~\Omega$ and $\alpha = 1.4$, where the $\ket{ee}$ state can be accessed with a speedup over the off-resonant static drive. The observed $\ket{ee}$ population is limited by the finite position spread of the atoms at about 1~$\mu$K, which gives rise to a range of interaction strengths, over which the resonance condition $\omega_0 = V$ does not always hold. This problem can be mitigated by optimal cooling techniques, such as motional ground-state cooling \cite{kaufman2012cooling}, and by increasing the atom-laser coherence (see Supplementary Information). 

To boost the $\ket{ee}$ population further, we propose to combine FFM with stimulated Raman adiabatic passage (STIRAP). A reliable state-preparation method, STIRAP has been used to initialize atoms in multiply-excited Rydberg states for studies of symmetry-protected topological phases \cite{de2019observation}, spin transport in one-dimensional systems \cite{scholl2022microwave}, and more. However, to date, such STIRAP transfer has only been demonstrated on atoms spaced outside the blockade radius. Reducing the spacing between atoms would be desired for stronger interactions, yet the steady-state population of multiply-excited Rydberg states cannot be achieved by applying STIRAP with the conventional static excitation scheme. 

On the other hand, FFM offers a straightforward path for populating the $\ket{ee}$ state through STIRAP for atoms well within the blockade radius. The modulation index is a flexible degree of freedom that controls both two-atom coupling strengths simultaneously. The adiabatic transfer of atoms from the initial $\ket{gg}$ state to the final $\ket{ee}$ state can be accomplished by ramping the modulation index from its first Bessel function zero ($\alpha$ = 2.4; $J_0(\alpha) = 0$) down to $\alpha = 0$ (where $J_1(\alpha) = 0$) over time (Fig.~4d). The adiabatic ramp needs to be performed quickly compared to the decoherence of the $\ket{W}$ state but slowly compared to the coupling strengths $\{\Omega_{a}, \Omega_{b}\}$. We note that the effective Rabi frequency for each transition varies asymmetrically in time despite the laser intensities being kept constant throughout the transfer.

%%%%%%%%%%%%%%%%%%%%%%%% OUTLOOK %%%%%%%%%%%%%%%%%%%%%%%%%%%%%
\paragraph*{Discussion and outlook.} We have demonstrated that FFM is a versatile approach that can be used to increase the entanglement range, protect the entangled state coherence, and initialize
strongly-interacting states in Rydberg atom arrays. The full advantages afforded by FFM can be realized by working with a Rydberg state of higher principal quantum number so as to access a longer lifetime and by making several technical upgrades. These include improving the Rydberg laser coherence though cavity filtering \cite{levine2018high}, suppressing off-resonant scattering rates by increasing the detuning from the intermediate state \cite{de2018analysis, evered2023high}, and reducing the position disorder through ground-state cooling \cite{kaufman2012cooling}, all of which have already been demonstrated separately in other experiment setups.

Our results can be extended to the generation and protection of entanglement between two long-lived ground states, where the entanglement is mediated by excitation to a Rydberg state \cite{levine2019parallel}. For an arbitrary graph depicting a particular geometric arrangement of single atoms at its vertices, FFM enables connectivity between any two atoms through an appropriate choice of the modulation index. FFM can also be combined with mobile optical tweezers that can transport entanglement over longer distances \cite{bluvstein2022quantum}. In such a combination, FFM can reduce both the duration and the number of moves required to perform pairwise entanglement across the entire array, thereby leading to a more streamlined quantum information processing architecture \cite{saffman2016quantum}. We further note that the enlarged blockade range can be used to entangle multiple atoms simultaneously (see Supplementary Information). In this case, FFM can effect the dynamical control of the Rydberg blockade range beyond that accomplished by simply tuning the Rydberg laser intensity, thereby offering a flexible way to access quench dynamics in quantum many-body simulations \cite{ebadi2021quantum, scholl2021quantum}.

Importantly, our work redefines the ability to access the two paradigmatic regimes of interactions --- Rydberg blockade versus anti-blockade, which have thus far been mostly governed by the precise positioning of atoms and the static blockade radius. The FFM not only enhances the programmability of Rydberg atom arrays but also enables steady-state Rydberg anti-blockade for atoms spaced within the blockade radius, which cannot be otherwise attained with conventional static schemes. These results open the door to realizing arrays of closely-spaced Rydberg atoms, including those in long-lived circular Rydberg states that are attractive for quantum computing and simulation \cite{nguyen2018towards, meinert2020indium, cohen2021quantum}.

%%%%%%%%%%%%%%%%%%%%%%%% METHODS %%%%%%%%%%%%%%%%%%%%%%%%%%%%%

\section*{Methods}
Our experiment begins with loading individual $^{23}$Na atoms into a pair of magic-wavelength tweezer sites from a pre-cooled reservoir \cite{aliyu2021d}. Each of the two traps is formed by focusing 615.87~nm light to a beam waist of 0.78~$\mu$m and is set to a trap depth of 1.3~mK during the $D_{1}$ $\Lambda$-enhanced loading. We then image once with $\Lambda$-enhanced $D_1$ imaging to post-select experiment shots in which both traps are loaded. Unlike in \cite{aliyu2021d}, here the counterpropagating pair of $D_1$ imaging beams are sent along the $y$ axis and are $(\sigma^{+}, \sigma^{-})$ polarized with a total intensity of 6~$I_\mathrm{sat}$ ($I_\mathrm{sat} = 6.26~$mW/cm$^{2}$) and $I_\mathrm{repump}/I_\mathrm{cool} = 0.11$.

We initialize the atoms in the ground state $\ket{g} = \ket{3 S_{1/2}, F = 2, m_{F} = 2}$ with 99.1\% fidelity by optically pumping the atoms in the presence of a bias magnetic field of 2.7~G along the $x$ axis. The trap depths are then ramped linearly down to 7~$\mu$K in 0.8~ms, yielding an atom temperature of about 1~$\mu$K as determined by comparing the measured release-and-recapture probability against Monte Carlo simulations. The tweezer traps are subsequently turned off during Rydberg excitation.

To drive the atoms to the $\ket{59 S_{1/2}, m_{J} = 1/2}$ Rydberg state, a global two-photon excitation pulse consisting of a $\sigma^{+}$-polarized 589~nm beam and a counter-propagating $\sigma^{-}$-polarized 409~nm beam is sent to the atoms along the $x$ axis. We realize single-photon Rabi frequencies of $\{\Omega_{1}$, $\Omega_{2}\}$ $= 2\pi \times \{70, 26\}$~MHz and single-photon detunings of $\Delta' \approx 2\pi \times 850$~MHz, red-detuned from the $\ket{3 P_{3/2}, F = 3, m_{F} = 3}$ intermediate state. 
We note that the atom array axis (along $y$) being orthogonal to the direction of Rydberg laser propagation (along $x$) is suboptimal, but this is limited in our experiment setup by the residual coma aberration of the microscope objective, and can be easily resolved with an improved objective. Nevertheless, we equalize the Rabi frequencies for the two atoms by realigning the 409~nm laser with a piezoelectric mirror from time to time. The combined two-photon Rydberg Rabi frequency is typically $\Omega = 2\pi \times 1.0$~MHz. Our atom array orientation also means that the relative phase $\phi$ in the expression for $\ket{W}$ is close to zero.

After the Rydberg excitation pulse, atoms in the ground (Rydberg) state will be recaptured (repelled) by the tweezer trap (restored trap depth of 1.35~mK). Finally, the $D_1$ imaging beams are applied a second time to read out the population of atoms in the ground versus Rydberg state. A bright (dark) image nominally indicates the presence of an atom in its ground state (Rydberg state). The data reported in the main text include the state preparation and measurement errors (see Supplementary Information). Each data point is the average of at least 200 experiment cycles.

%%%%%%%%%%%%%%%%%%%%%%%% DATA AND CODE AVAILABILITY %%%%%%%%%%%%%%%%%%%%%%%%%%%%%

\section*{Data availability} The data that support the findings of this study are available on \href{https://doi.org/10.5281/zenodo.8347276}{https://doi.org/10.5281/zenodo.8347276}.

\section*{Acknowledgements}
We acknowledge Travis Nicholson and Wen Wei Ho for stimulating discussions, as well as Krishna Chaitanya Yellapragada for technical assistance with the experiment setup. This research is supported by the National Research Foundation, Singapore and A*STAR under its Quantum Engineering Programme (NRF2021-QEP2-02-P09) and its CQT bridging grant. 

%\section*{Code availability} The simulation codes used in this work are available from the corresponding author upon request.

%%%%%%%%%%%%%%%%%%%%%%%%%% REFS AND ACKNOWLEDGMENTS %%%%%%%%%%%%%%%%%%%%%%%%%%%%%%

%%%%%%%%%%%%%%%%%%%%%%%%%%%%%%%%%%%%%%%%%%%%%%%%%%%%%%%%%%%%%%%%%%%%%%%%%%%%%%%%%%%%%%%
%\bibliographystyle{sn-nature}
%\bibliography{references}

%\def\bibsection{\section*{\refname}}

%%%%%%%%%%%%%%%%%%%%%%%%%%%%%%%%%%%%%%%%%%%%%%%%%%%%%%%%%%%%%%%%%%%%%%%%%%%%%%%%%%%%%%%

\clearpage
\setcounter{equation}{0}
\setcounter{figure}{0}
\setcounter{table}{0}
\renewcommand{\figurename}{Supplementary Figure}
\onecolumngrid

\begin{center}
\large{\textbf{Supplementary Information: Floquet-Tailored Rydberg Interactions}} \\
\end{center}
\begin{center}
Luheng Zhao,$^{1}$ Michael Dao Kang Lee,$^{1}$ Mohammad Mujahid Aliyu,$^{1}$ and Huanqian Loh$^{1,2,\ast}$\\ \vspace{1mm}
\small{\textit{$^{1}$Centre for Quantum Technologies, National University of Singapore, 117543 Singapore, Singapore}\\
\textit{$^{2}$Department of Physics, National University of Singapore, 117542 Singapore, Singapore}}\\
\end{center}

\twocolumngrid
\section{State preparation and measurement errors} \label{sec:SPAM}
The ground and Rydberg states are distinguished by the atom survival and loss after the second imaging pulse is applied during the experiment sequence. However, the probabilities of detecting the ground and Rydberg states ($P_{g}, P_{e}$) are susceptible to false positive errors $\epsilon$ and false negative errors $\epsilon'$, which can obscure the true probabilities $(\tilde{P}_{g}, \tilde{P}_{e})$.

False positive errors $\epsilon = P(e|\tilde{g})$ refer to the case where lost ground state atoms are misinterpreted as Rydberg atoms. The loss of ground state atoms can occur due to background collisions, the atom-recapture inefficiency, and the heating of ground state atoms during imaging. In our experiment, the last mechanism dominates the false positive error, which is measured to be 0.03. We note in passing that while the $D_1$ imaging survival probability in this experiment is 0.97, it can generally be higher (0.99) if the imaging configuration previously reported in \cite{aliyu2021d} is used. 

On the other hand, false negative errors $\epsilon' = P(g|\tilde{e})$ arise from Rydberg atoms that have decayed to the ground state and are recaptured before exiting the tweezer capture range. These atoms are consequently miscounted as ground state atoms. In contrast, atoms that remain in the Rydberg state would be repelled by the anti-trapping potential of the tweezer. For a 1.3~mK recapture trap depth, the estimated time required for Rydberg atoms to leave the trapping region is 4.5~$\mu$s. Given the natural lifetime of 260~$\mu$s for the $59 S_{1/2}$ Rydberg state \cite{beterov2009quasiclassical}, we expect a false negative error of 0.02. In addition, before the tweezer light is turned back on, there is a delay of 0.5~$\mu$s during which only the 409~nm Rydberg laser is left on. Combined with the off-resonant scattering rate of the 409~nm laser, this delay adds 0.01 to the false negative error. Therefore the total false negative error is 0.03.

The detected and true populations are then related to each other by the following expressions \cite{de2018analysis}:
\begin{equation}
    P_{g} = \eta (1-\epsilon) + (1-\eta) (1 - \epsilon) (\tilde{P}_{g} + \epsilon' \tilde{P}_{r}) \\ ,
\end{equation}
\begin{equation}
    P_{r} = \eta \epsilon + (1-\eta) (\epsilon \tilde{P}_{g} + (1-\epsilon' +\epsilon \epsilon') \tilde{P}_{r}) \, ,
\end{equation}
where $\eta$ is the error associated with imperfect optical pumping and is estimated to be 0.009 in our experiment. For the two-atom populations, the above probabilities are multiplied accordingly.

%%%%%%%%%%%%%%%%%%%%%%%%%%%%%%%%%%%%%%%%%%%%%%%%%%%%%%%%%%%%%%%%%%%%%%%%%%%%%%%%%%
\section{Calibration of atom separation}
\begin{figure}[!htbp]
	\centering
	\includegraphics[keepaspectratio,width=8.2cm]{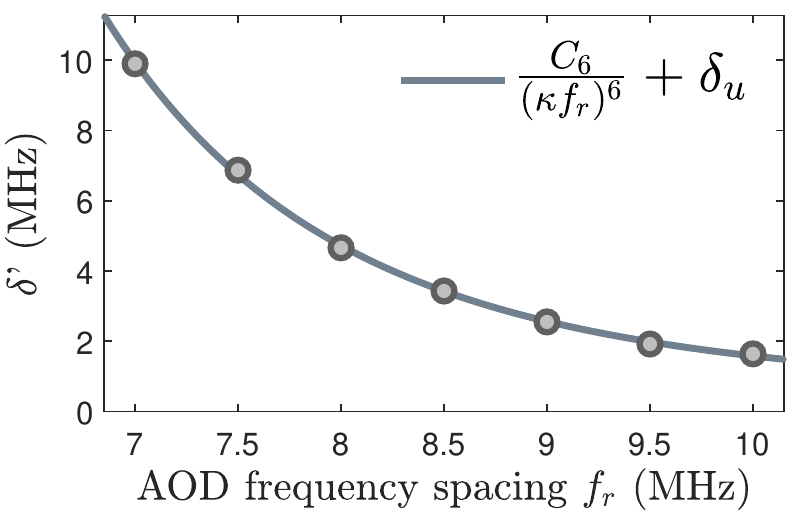}
        \caption{Calibration of the interatomic distance $r$ with respect to the AOD frequency spacing $f_{r}$. Error bars are smaller than the displayed marker sizes. The calibration factor $\kappa$ is determined to be 0.780(2)~$\mu$m/MHz.} 
	\label{fig:calibAOD}
\end{figure}

Precise control of the separation distance between atoms is important as it determines the strength of interaction between the atoms when excited to Rydberg states. In the FFM studies, the nominal atom spacing ranges from 5.6~$\mu$m to 10.3~$\mu$m. The spacing is controlled by the frequencies sent to the acousto-optic deflector (AOD). Calibration of the AOD frequency spacing is performed by directly measuring the interaction-induced energy shift experienced by two atoms in the $\ket{ee}$ state with the following sequence: first, the two atoms are excited from $\ket{gg}$ to $\ket{W}$ with a resonant monochromatic $\pi$-pulse ($\delta = 0$);  subsequently, $\ket{W}$ is excited to $\ket{ee}$ with a second monochromatic pulse of variable detuning $\delta'$. The $\ket{ee}$ population is maximized when $\delta' = V$. The above sequence is repeated at different AOD frequency spacings. The measurements are fit with the calculated value \cite{arc} of $C_{6}$ for the $59 S_{1/2}$ Rydberg state of $^{23}$Na, $C_{6} = 2\pi \times 251.288~$GHz $\mu$m$^{6}$, and an overall distance-independent frequency offset $\delta_{u}$ arising from an imperfection in determining the unshifted resonance (Supplementary Fig.~\ref{fig:calibAOD}). The fit distance calibration of 0.780(2)~$\mu$m/MHz is consistent with the expected value (0.783~$\mu$m/MHz) calculated from the acousto-optic deflector specifications, objective focal length, and tweezer telescope magnification.

%%%%%%%%%%%%%%%%%%%%%%%%%%%%%%%%%%%%%%%%%%%%%%%%%%%%%%%%%%%%%%%%%%%%%%%%%%%%%%%%%%
\section{FFM details}
The FFM is implemented by sending the 589~nm Rydberg laser through an acousto-optic modulator (AOM), which is driven by an arbitrary waveform generator (AWG) to yield a time-varying frequency (-$\delta \sin \omega_{0}t$). We ensure that the frequency-modulated RF signal has a well-defined phase by triggering the AWG. The AOM is used in a double-pass configuration so that the output beam angle sent to the neutral atom array remains unchanged despite the time-varying single-pass diffraction angle. To enable a fast on-off response of the AOM, we focus the 589 nm Rydberg excitation laser to a beam waist of 35~$\mu$m in the AOM. 

\subsection{Calibration of modulation index}
Since the dynamics of the two-atom system depends on the modulation index $\alpha$, it is important to calibrate the modulation amplitude $\delta$, which can differ from the AWG-specified amplitude due to the finite AOM bandwidth. This is performed by monitoring an optical beat note between the first-order deflected beam with a reference beam. For a given applied modulation frequency $\omega_{0}$, we measure the power of the beat note carrier while changing the modulation amplitude $\delta$. The normalized carrier power measured with respect to the modulation index $\alpha$ is fitted to the modulus of the zeroth order Bessel function with a free linear scaling factor $\chi$ (Supplementary Fig.~\ref{fig:calibffm}a, b). We find that the AWG-specified modulation index is slightly larger than the theoretical value by $\chi = 1.009(2)$ and 1.0450(9) for $\omega_0$ = 3 MHz and 6 MHz, respectively. This means that the actual modulation index effected by the AOM is correspondingly smaller than the AWG-specified value. We note that $\chi$ is sensitive to the laser beam alignment through the AOM.

To further investigate the effect of the modulation frequency on $\chi$, we measure the first Bessel function zero $\alpha_0$ (i.e.\ where $J_0(\alpha_0) = 0$) for different modulation frequencies by tracking the modulation amplitude at which the carrier power is minimized. Similarly, we observe the measured modulation index to be higher than the theoretical Bessel zero of 2.40. The finite AOM bandwidth yields a larger deviation for higher modulation frequencies, as shown in Supplementary Fig.~\ref{fig:calibffm}c.

\subsection{Minimization of residual amplitude modulation}
Having a residual amplitude modulation (RAM) is undesirable as it would change the dynamics from that predicted by the Hamiltonian given in:
\begin{equation} \label{eq:Hnative}
\frac{H}{\hbar} = -\Delta(t) \sum_{i=1}^{N} \sigma_{ee}^{i} + \frac{\Omega}{2}\sum_{i=1}^{N} \sigma_{x}^{i} + \sum_{i<j} V_{ij} \sigma_{ee}^{i} \sigma_{ee}^{j} \, ,
\end{equation}
In particular, the entanglement coherence will be impacted most significantly as the dynamical freezing of $\ket{W}$ is sensitive to the ability to precisely zero the two-atom Rabi frequency $\Omega_{a}$. We model the RAM as a modification of the Rabi frequency between the $\ket{gg}$ and $\ket{W}$ states from
\begin{equation} 
\label{eq:omegaa} 
\Omega_{a}(t) \propto \Omega\sum_{m=-\infty}^{\infty} J_{m}(\alpha) e^{i m \omega_0t + im\frac{\pi}{2}} \,
\end{equation}
to
\begin{widetext}
\begin{eqnarray}
\Omega_{a} &\rightarrow& \Omega \left[1 + \sum_{n=1}^{\infty} (A_{n} \sin n \omega_{0} t + B_{n} \cos n\omega_{0} t)\right] \sum_{m=-\infty}^{\infty} J_{m} (\alpha) e^{im\omega_{0}t + im \frac{\pi}{2}} \, \nonumber \\
&=& \Omega \left[1 + \sum_{n=1}^{\infty} (A_{n} \sin n \omega_{0} t + B_{n} \cos n\omega_{0} t)\right] \left[J_{0}(\alpha) + \sum_{m=1}^{\infty} J_{m}(\alpha) e^{im\frac{\pi}{2}} \cos m\omega_{0} t\right] \hspace{5mm}
\label{eqn:ram}
\end{eqnarray}
\end{widetext}
where $A_{n}$ and $B_{n}$ are assumed to be small. Eq.~(\ref{eqn:ram}) shows that to lowest order, only the cosine components of the RAM would perturb $J_{0}(\alpha)\Omega$, whereas the sine components of the RAM (i.e.\ components in phase with FFM) exert minimal effect on $\Omega_{a}$. 

The RAM can arise either from our modulation of the single-photon detuning $\Delta'$, which leads to an effective modulation of the two-photon Rabi frequency $\Omega$, or from the frequency-dependent diffraction efficiency of the AOM. For the former, we estimate an effective modulation of  $\delta/\Delta' = 4$\% on $\Omega$ for the experiments involving the largest modulation indices. However, since the modulation on $\Omega$ is in phase with the FFM, we did not attempt to compensate for this factor. On the other hand, the RAM arising from the AOM imperfection is minimized in our experiment by either calibrating the frequency-dependent diffraction efficiency or combining the calibration with negative feedback based on gradient descent. 

\begin{figure}[!htbp]
	\centering
	\includegraphics[keepaspectratio,width=8.2cm]{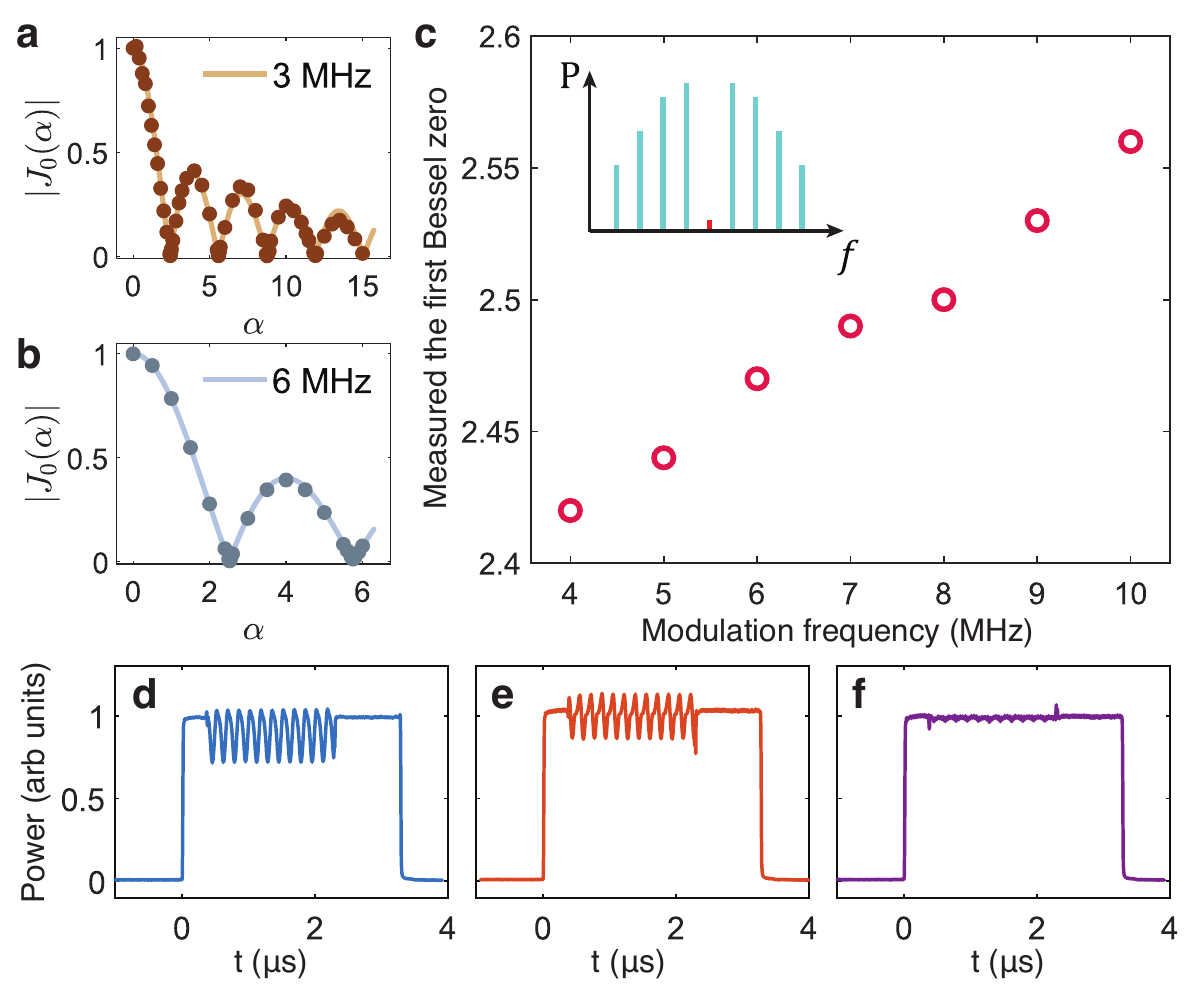}
        \caption{Mitigation of FFM imperfections.  \textbf{a, b} Calibration of modulation index for \textbf{a} $\omega_{0} = 3~$MHz and \textbf{b} $\omega_{0} = 6~$MHz, by measuring the normalized carrier power versus AWG-specified modulation index $\alpha$. The measured data is fit to $|J_{0}(\alpha)|$. \textbf{c} Calibration of modulation index for $\omega_{0} = 4-10$~MHz, performed by measuring the positions of the Bessel function zeros $\alpha_{0}$, where $J_0(\alpha_{0}) = 0$. (Inset) Schematic of the optical beat note power spectrum, where the (red) carrier is minimized at $\alpha_{0}$. \textbf{d} Measured Rydberg laser excitation pulse with FFM ($\omega_{0} = 6~$MHz, $\alpha = 5.5$) and no amplitude compensation. \textbf{e} Measured Rydberg laser excitation pulse with the same FFM parameters, incorporating a calibration of the AOM frequency-dependent diffraction efficiency. \textbf{f} Measured Rydberg laser excitation pulse with same FFM parameters and both calibration and gradient descent compensation.} 
	\label{fig:calibffm}
\end{figure}

For the calibration, a quadratic correction factor was initially applied to the AOM: 
\begin{equation}
    V_\mathrm{AOM} \to \frac{V_\mathrm{AOM}}{a(f-f_c)^2 + c} \, ,
\end{equation}
where the coefficients $a, f_{c}$, and $c$ were obtained by minimizing the power variation as the frequency of the AOM was linearly ramped. We then conducted a more detailed calibration of the AOM diffraction efficiency by fitting the output power to the following function:
\begin{equation}
    P(f,V_\mathrm{AOM}) = A(f) \tanh\left(\frac{V_\mathrm{AOM}-V_0(f)}{\sigma(f)}\right) + C(f) \, ,
\end{equation}
where the frequency-dependent fitting parameters $A(f)$, $V_{0}(f)$, $\sigma(f)$, and $C(f)$ were in turn determined by a polynomial fit to the frequency. To obtain a constant power output $P$ independent of $f$, we invert the expression to get the AOM amplitude required:
\begin{equation}
    V_\mathrm{AOM}(P,f) = V_{0}(f) + \sigma \tanh^{-1} \left(\frac{P-C(f)}{A(f)}\right) \, .
\end{equation}
Through this detailed calibration, we can reduce the power output error from $\approx$30\% peak-to-peak variation to $<$10\% peak-to-peak variation when we manually change the frequency tones over a 50 MHz bandwidth. However, when we perform FFM, we observe that the calibration only reduces the DC offset error (Supplementary Fig.~\ref{fig:calibffm}e), while there is still an error that is synchronous with the frequency modulation.

To further mitigate the RAM, we perform an additional feedback process. We iteratively add a set of $n$ harmonics of the modulation frequency, where $n = 1$ -- 4 with both quadratures (sine and cosine) applied. The amplitudes of these harmonics are varied using gradient descent. Supplementary Fig.~\ref{fig:calibffm}f shows a minimized RAM of 2.2\% when both calibration and gradient descent optimization methods are employed.

%%%%%%%%%%%%%%%%%%%%%%%%%%%%%%%%%%%%%%%%%%%%%%%%%%%%%%%%%%%%%%%%%%%%%%%%%%%%%%%%%%
\section{Numerical simulation of dynamics} \label{sec:numerics}
For our numerical modeling, we use the Python package QuTiP \cite{qutip} to simulate the dynamics as governed by the Lindblad master equation \cite{levine2018high, de2018analysis, tamura2020analysis}:
\begin{equation}
    \dot{\rho} = -\frac{i}{\hbar}[H, \rho] + \mathcal{L}[\rho] \, .
\end{equation}
$\mathcal{L}[\rho]$ is a sum of the following three terms corresponding to off-resonant scattering from the intermediate state $\mathcal{L}_{m}$, spontaneous decay from the Rydberg state $\mathcal{L}_{r}$, and laser phase noise $\mathcal{L}_{l}$, respectively:
\begin{equation}
    \mathcal{L}_{m}[\rho]= \sum_{i}(L_{m}^{i} \rho (L_{m}^{i})^{\dagger} - \frac{1}{2}\{(L_{m}^{i})^{\dagger} L_{m}^{i}, \rho\}) \, ,
\end{equation}
\begin{equation}
    \mathcal{L}_{r}[\rho]= \sum_{i}(L_{r}^{i} \rho (L_{r}^{i})^{\dagger} - \frac{1}{2}\{(L_{r}^{i})^{\dagger} L_{r}^{i}, \rho\}) \, ,
\end{equation}
\begin{equation}
    \mathcal{L}_{l}[\rho]= L_{l} \rho L_{l}^{\dagger} - \frac{1}{2}\{L_{l}^{\dagger} L_{l}, \rho\} \, ,
\end{equation}
where $i$ indexes the atom. 

The off-resonant scattering can arise from both the 589~nm Rydberg laser $\gamma_{1}$ and the 409~nm Rydberg laser $\gamma_{2}$, which are described by their corresponding Lindblad operators $L_m^{(i)} $ $= $ $\sqrt{\gamma_{1}} |g_{i}\rangle \langle g_{i}| + \sqrt{\gamma_{2}} |g_{i}\rangle \langle e_{i}|$. In the experiment, $\gamma_{1} \approx 2\pi \times 17$~kHz and $\gamma_{2} \approx 2\pi \times 2.4$~kHz. 
%Table~\ref{tab:exptparams} gives the off-resonant scattering rates for the different data sets reported in the main text. 
The blackbody-radiation-limited lifetime \cite{beterov2009quasiclassical} of the $59 S_{1/2}$ Rydberg state is $1/\Gamma_{r} = $ 106.5~$\mu$s and is modeled with the simplified Lindblad operator $L_{r}^{(i)} = \sqrt{\Gamma_{r}}|g_{i}\rangle \langle e_{i}|$. Finally, the laser phase noise $\gamma_{l}$ is modeled as a global dephasing term: $L_{l} = \sqrt{\gamma_{l}/2}\, \sigma_{z}^{(1)} \otimes \sigma_z^{(2)}$. $\gamma_{l}$ is inferred from the coherence time of the single-atom Rabi oscillation between $\ket{g}$ and $\ket{e}$ to be $2\pi \times 50$~kHz. 

The calculated populations are then scaled to include the state preparation and measurement errors described in Section~\ref{sec:SPAM}.

\begin{figure}[!tbhp]
	\centering
	\includegraphics[keepaspectratio,width=8.2cm]{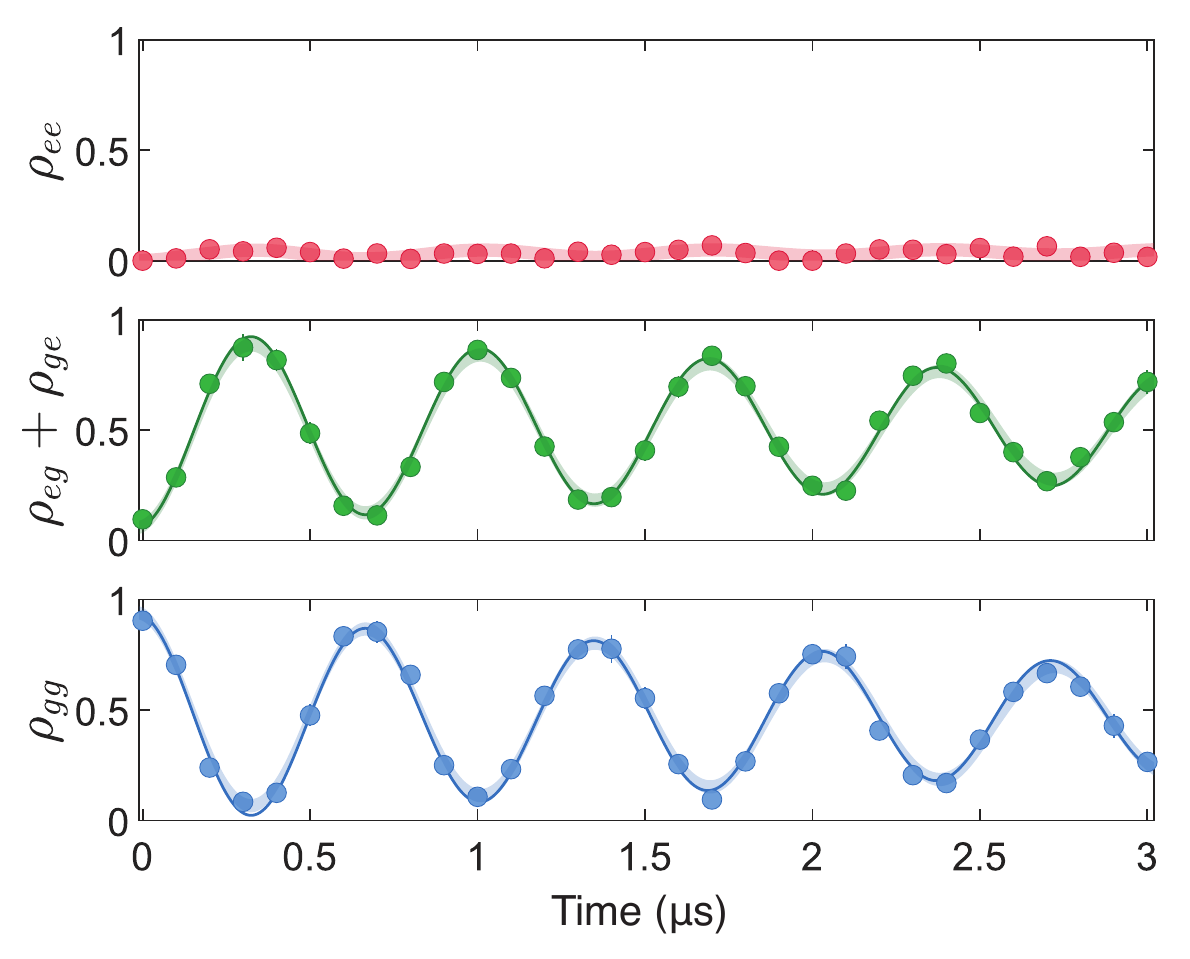}
        \caption{Two-atom dynamics under static Rydberg excitation with $V = 6~\Omega$ for (top) $\rho_{ee}$, (middle) $\rho_{eg} + \rho_{ge}$, and (bottom) $\rho_{gg}$. The markers indicate the experimental measurements, which are fit (solid line) to yield a collective Rabi frequency of $2\pi \times 1.466(3)$~MHz and a 5.0(6)~$\mu$s decay time constant. The shading reflects the Monte Carlo simulation results.}
	\label{fig:twoatom}
\end{figure}

%Due to their finite temperature, the atoms have an initial radial and axial position spread in the tweezers, 
At a temperature of 1.2~$\mu$K, the atoms have a finite radial and axial position spread of $\{\sigma_{x, y}, \sigma_{z}\}$ $\approx \{0.17~\mu$m, $0.92~\mu$m$\}$ in the tweezers, 
which increase slightly when the tweezers are turned off during the Rydberg excitation. We model this effect by using Monte Carlo methods to randomly sample the atom position from a normal distribution determined by its initial position uncertainty. We then take into account the effect of the finite velocity, which changes the atom spacing over time. The Doppler shifts for each atoms are also randomly sampled from normal distributions. The simulated two-atom populations are sorted into histograms, of which the standard deviations are depicted as shaded curves in Figs.~2c inset, 2d, 2e, 3b, 3c, and 4c of the main text. In Fig.~2c (main plot) of the main text, the Monte Carlo sampling is not depicted for clarity. As a sanity check, Supplementary Fig.~\ref{fig:twoatom} shows the observed two-atom dynamics driven by a static Rydberg excitation scheme, where $V = 6~\Omega$ and where the measured populations are in good agreement with those from the Monte Carlo simulations with the above Lindblad terms.

In Fig.~4 of the main text, we note that the $\ket{ee}$ population at short times (Fig.~4c) can be boosted by using a less aggressive ramp of the trap depth, since the interaction disorder there mainly comes from the initial position spread. Where we consider the anti-blockade dynamics under ground-state cooling, we assume a final motional state occupation of $\{\bar{n}_{x, y}, \bar{n}_{z}\} = \{0.05, 0.20\}$. We further assume that the atoms can remain trapped during the Rydberg excitation, such that there is no further increase in the position uncertainty due to time-of-flight dynamics. For the idealized scenario combining ground-state cooling with enhanced coherence (Figs.~4c, d of the main text), we assume suppressed off-resonant scattering rates and a phase noise of $2\pi \times 1$~kHz for each laser. In addition, we assume working in the $n = 80$ Rydberg state of $^{23}$Na, which offers a longer lifetime\cite{beterov2009quasiclassical}. The combined suppression of decoherence effects would increase the effective two-atom coherence time to 74~$\mu$s. For these idealized cases, the corresponding plots in Figs.~4c and 4d of the main text do not include state preparation and measurement (SPAM) errors.

%%%%%%%%%%%%%%%%%%%%%%%%%%%%%%%%%%%%%%%%%%%%%%%%%%%%%%%%%%%%%%%%%%%%%%%%%%%%%%%%%%
\section{Fidelity estimation}
\begin{figure}[!htbp]
	\centering
	\includegraphics[keepaspectratio,width=8.2cm]{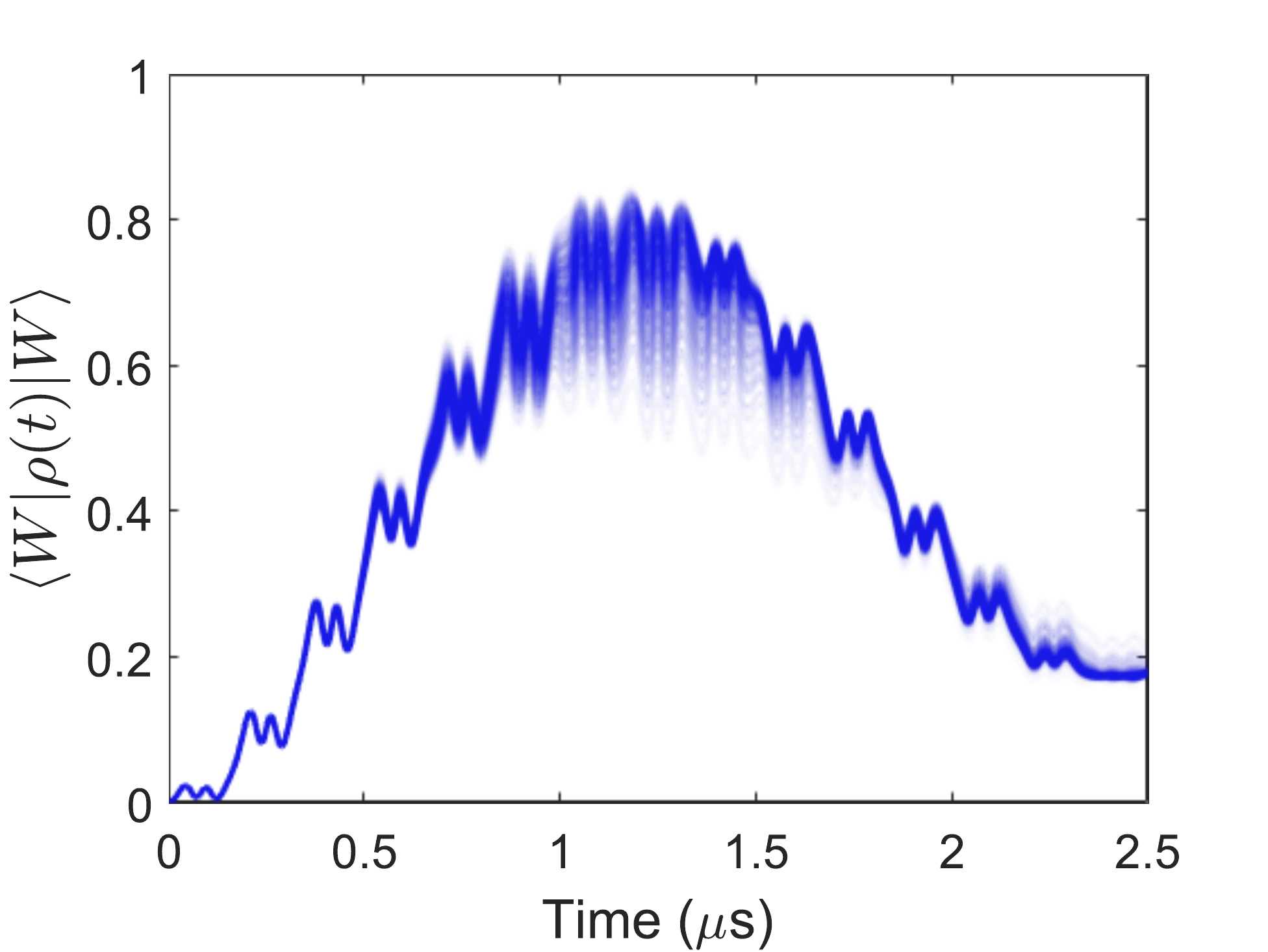}
        \caption{Calculated fidelities $\langle W | \rho(t) | W \rangle$. The time evolution of the density matrix $\rho(t)$ is obtained from numerically modeling the data shown in Fig.~2e of the main text, where $V = 0.8~\Omega, \omega_{0} = 3~\Omega$, and $\alpha = 6.9$. The different curves indicate various initial atom positions and velocities randomly sampled from normal distributions, whose standard deviations correspond to an atom temperature of 1.2~$\mu$K. }
	\label{fig:fidelity}
\end{figure}

To estimate the fidelity of the $\ket{W}$ state generated in the enhanced blockade regime, we first solve the Lindblad master equation over the excitation time $t$ to yield the density matrix $\rho^{(j)}(t)$ for a given set of random initial atom positions and velocities. We use the same initial positions $\{x_{1}^{(j)}, x_{2}^{(j)}\}$ to define the symmetric state \cite{levine2018high}:
\begin{equation}
    \ket{W^{(j)}} = e^{i k x_{1}^{(j)}} \left(\ket{ge} + e^{i k (x_{2}^{(j)}-x_{1}^{(j)})} \ket{eg} \right) \, ,
\end{equation}
where $j$ indexes the set of initial conditions and $k$ is the effective wavevector of the two counterpropagating Rydberg lasers. We note that the relative phase $k (x_{2}^{(j)}-x_{1}^{(j)})$ is nominally fixed for each experiment trial. We calculate the fidelity $\mathcal{F}^{(j)}(t) = \langle W^{(j)} | \rho^{(j)}(t) | W^{(j)} \rangle$ as a function of time, while taking note of the maximum fidelity $\mathcal{F}_\mathrm{max}^{(j)}$ achieved during the excitation period. The above procedure is repeated for different sets of initial conditions randomly sampled from their respective normal distributions as described in Section~\ref{sec:numerics}. Supplementary Fig.~\ref{fig:fidelity} shows the calculated fidelities $\mathcal{F}^{(j)}(t)$ for different initial conditions corresponding to the data shown in Fig.~2e of the main text. The reported fidelity of 0.77(5) is the mean and standard deviation of $\mathcal{F}_\mathrm{max}$.

%%%%%%%%%%%%%%%%%%%%%%%%%%%%%%%%%%%%%%%%%%%%%%%%%%%%%%%%%%%%%%%%%%%%%%%%%%%%%%%%%%
\section{Enhanced qubit connectivity}
The $\ket{W}$ fidelity of 0.98 reported in the main text is calculated for $\omega_{0} = 3~\Omega$, $\alpha = 11.1$, and $V = 0.5~\Omega$ in the absence of decoherence and SPAM errors. For a static drive, the atoms would need an interaction strength of at least $V = 4.9~\Omega$ to achieve the same fidelity. 

\begin{figure}[!htbp]
	\centering
	\includegraphics[keepaspectratio,width=8.2cm]{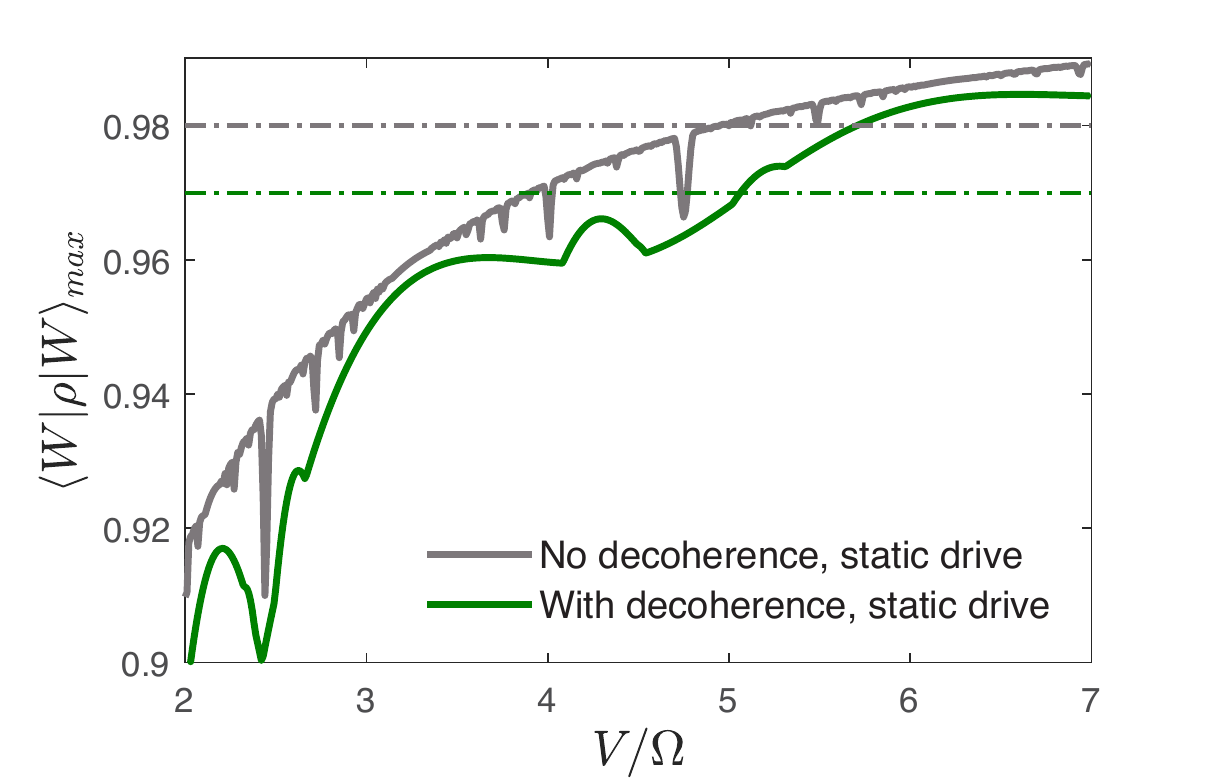}
        \caption{Comparison of maximum $\ket{W}$ fidelities achieved in a static Rydberg excitation against that from FFM. Assuming no decoherence effects, FFM ($\omega_{0} = 3~\Omega$, $\alpha = 11.1$, $V = 0.5~\Omega$) can yield (gray dot-dashed line) a fidelity of $\langle W | \rho | W \rangle_\mathrm{max} = 0.98$. To obtain the same fidelity in a static drive, atoms have to spaced more closely such that they experience (gray curve) $V = 4.9~\Omega$. With a finite effective coherence time of 74~$\mu$s, the FFM fidelity decreases to (green dot-dashed line) 0.97. However, the atoms still have to experience an interaction strength of at least (green curve) $V = 4.3~\Omega$ for a static drive to achieve the same fidelity.}
	\label{fig:qubitconnectivity}
\end{figure}

Assuming a finite effective coherence time of 74~$\mu$s, the $\ket{W}$ fidelity under FFM reduces to 0.97 for $V = 0.5~\Omega$. In this case, the corresponding interaction strength would have to be $V = 4.3~\Omega$ for atoms under the static Rydberg excitation scheme to experience the same $\ket{W}$ fidelity  (Supplementary Fig.~\ref{fig:qubitconnectivity}). Nevertheless, since $(4.3/0.5)^{1/6} > \sqrt{2}$, it remains that for a square array, where nearest neighbors can be pairwise entangled using the static scheme, an atom can be pairwise entangled with its next-nearest neighbor with the same fidelity when driven by FFM. 

We note that the same effective extension of the Rydberg blockade range can be achieved in the static scheme with a lower Rabi frequency $\Omega_{s}$, where $\Omega_{s} = \Omega J_{0}(\alpha)$. However, dynamic control of the Rydberg blockade range through FFM can potentially be simpler than that in the static scheme, especially where the atoms require two-color Rydberg excitation. In the latter case, accurately controlling a range of Rabi frequencies would require calibrating out the change in differential light shift that accompanies the nominal change in Rydberg laser intensities. Achieving the desired Rydberg laser intensities in turn requires a careful characterization of the AOM diffraction efficiency, which depends on both its RF drive frequency and amplitude. In contrast, FFM only requires the latter characterization of the AOM to realize a drive with minimal residual amplitude modulation. Further, the control of Rabi frequencies over a large range (0 to $\Omega$) can already be achieved with a fairly small range of modulation indices (\textit{e.g.}\ $\alpha$ = 2.4 to 0).

\begin{figure}[t]
	\centering
	\includegraphics[keepaspectratio,width=8.2cm]{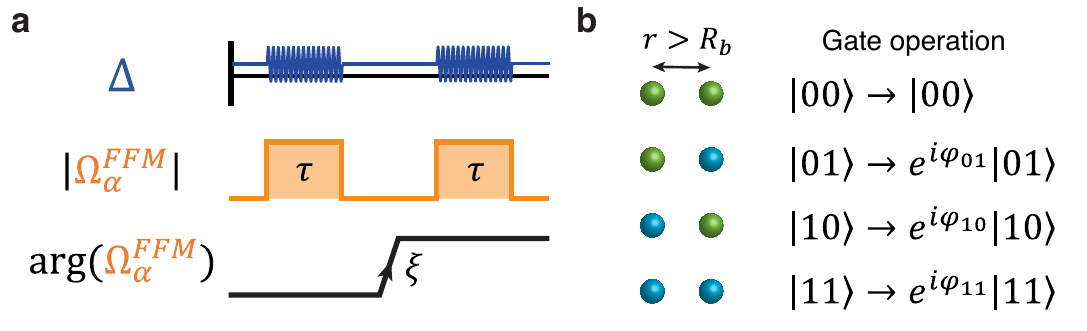}
        \caption{FFM-modified Levine-Pichler protocol \cite{levine2019parallel} for implementing a controlled-phase gate. \textbf{a} Where an extended Rydberg blockade range is desired, the two global Rydberg pulses of duration $\tau$ are applied with a frequency-modulated detuning given by $\Delta(t) = \Delta_{0} + \delta \sin (\omega_{0} t)$. \textbf{b} Even when two atoms are spaced farther than the blockade radius, the modified protocol can enable the two-qubit state to pick up a phase unless both qubits are in state $\ket{0}$.}
	\label{fig:LPFFM}
\end{figure}

While this work focuses on using the ground state $\ket{g}$ and Rydberg state $\ket{r}$ as the two qubit states, one can apply the FFM scheme to entangle atoms that encode their qubit states in two (\textit{e.g.}\ hyperfine) ground states, denoted here as $\ket{0}$ and $\ket{1}$. For instance, FFM is readily compatible with the Levine-Pichler protocol \cite{levine2019parallel} that has been used to implement a two-qubit controlled-phase gate in the Rydberg blockade regime \cite{jaksch2000fast}. The conventional Levine-Pichler protocol uses two global Rydberg pulses of duration $\tau$, sandwiched by a phase jump, to drive partial or full Rabi oscillations on two-qubit states with at least one qubit in $\ket{1}$. Where an extended Rydberg blockade range is desired, the global Rydberg pulses can be applied with a frequency-modulated detuning and with a user-defined modulation index $\alpha$ (Supplementary Fig.~\ref{fig:LPFFM}). We note, however, that as long as the interaction strength $V$ is decreased, the two-qubit gate fidelity will be compromised, as shown by the following expression for the intrinsic two-qubit gate error \cite{saffman2016quantum} $E_{min}$:
\begin{equation}
E_{min} = \frac{3(7\pi)^{2/3}}{8} \frac{1}{(V \tau_{R})^{2/3}}
\end{equation}
where $\tau_{R}$ is the Rydberg state lifetime.

%%%%%%%%%%%%%%%%%%%%%%%%%%%%%%%%%%%%%%%%%%%%%%%%%%%%%%%%%%%%%%%%%%%%%%%%%%%%%%%%%%
\section{Entanglement of multiple atoms}

\begin{figure}[!htbp]
	\centering
	\includegraphics[keepaspectratio,width=8.2cm]{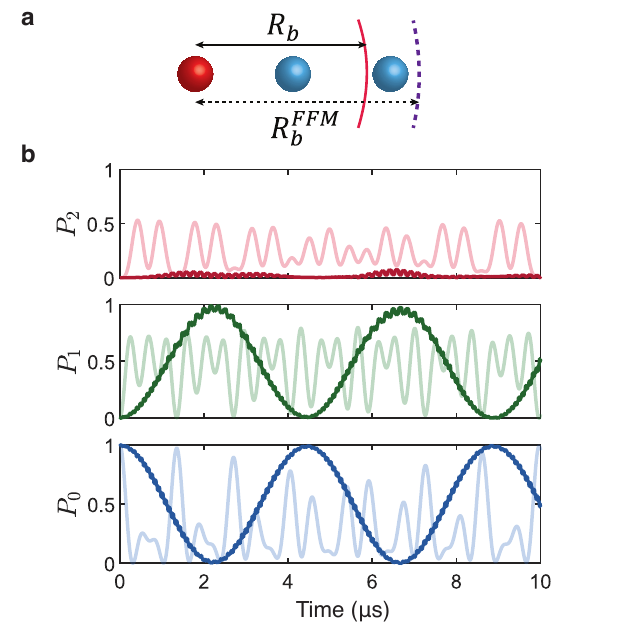}
        \caption{Increased entanglement range for multiple atoms.  \textbf{a} Three atoms are equally spaced in a chain. The interaction strength between next-nearest neighbors is $V = 0.5~\Omega$, as indicated by the static blockade radius $R_{b}$ (red curve). FFM is applied to extend the entanglement range to encompass next-nearest neighbors, as indicated by the extended blockade radius $R_{b}^\mathrm{FFM}$ (purple dashed curve). \textbf{b} Dynamics of three interacting atoms. $P_{n}$ indicates the probability of exciting $n$ atoms. For instance, $P_{1}$ maps onto the fidelity of the $\ket{W}$ state. (Solid curves) Under FFM ($\omega_{0} = 6~\Omega, \alpha = 11.2$), the three atoms can be driven to the entangled $\ket{W}$ state with high fidelity, where $P_{1} = 0.99$. Such a high fidelity cannot be otherwise achieved with (light curves) a static drive. }
	\label{fig:multipleatoms}
\end{figure}

For weak interactions $V < \Omega$, FFM can increase the effective Rydberg blockade radius relative to the static value, i.e. $R_{b}^\mathrm{FFM} > R_{b}$. Therefore, multiple atoms that reside within the increased blockade radius $R_{b}^\mathrm{FFM}$ can be entangled by driving them to the symmetric $\ket{W}$ state:
\begin{equation}
    \ket{W} = \frac{1}{\sqrt{N}} \left(\ket{e g g \ldots g} + \ket{g e g \ldots g} + \ket{g g e \ldots g} + \ldots \right)
\end{equation}
As an example, we consider three atoms equally spaced on a line, where one of the atoms resides outside the static blockade radius of the first atom (Supplementary Fig.~\ref{fig:multipleatoms}). Without FFM, the fidelity of the $\ket{W}$ state cannot exceed 0.80 for three atoms. However, with FFM, the fidelity is boosted to 0.99.

%%%%%%%%%%%%%%%%%%%%%%%%%%%%%%%%%%%%%%%%%%%%%%%%%%%%%%%%%%%%%%%%%%%%%%%%%%%%%%%%%%
\section{Antiblockade through STIRAP}

The steady-state population $\ket{ee}$ as shown in Fig.~4d of the main text can be achieved through the following STIRAP $\tanh$ profile:
\begin{equation}
    \frac{\alpha(t)}{\alpha_{0}} = 1.2\cdot\tanh\left[-\frac{3.5}{T}\left(t-\frac{T}{2}\right)+0.23 \right] + 1 \, ,
    \label{eqn:stirap}
\end{equation}
where $\alpha_{0} = 2.4$, such that $J_{0}(\alpha_{0}) = 0$, and $T$ is the duration of the STIRAP sequence. The initial condition allows us to minimize the $\ket{gg} \leftrightarrow \ket{W}$ Rabi frequency while yielding a reasonable $\ket{W} \leftrightarrow \ket{ee}$ coupling strength, whereas the final choice of $\alpha$ minimizes the $\ket{W} \leftrightarrow \ket{ee}$ Rabi frequency while giving the maximum $\ket{gg} \leftrightarrow \ket{W}$ coupling strength. The constants that appear in Eq.~(\ref{eqn:stirap}) are chosen such that $J_{0}(\alpha(t = 0)) = 0$, $J_{1}(\alpha(t = T)) = 0$  and $J_{0}(\alpha(t = T/2)) = J_{1}(\alpha(t = T/2))$. 

We note that this STIRAP approach is generally robust against timing imperfections. Nevertheless, since the time required to do the STIRAP transfer is 4~$\mu$s and is comparable to our present coherence times, we did not attempt a demonstration of our proposed STIRAP scheme here.

We have pointed out in the main text that it is not possible to populate the $\ket{ee}$ state for two closely spaced atoms via STIRAP with the conventional monochromatic static excitation scheme. That said, one can adopt a version of STIRAP with two static frequency components. In this case, where each single-atom Rydberg excitation requires two colors, the bichromatic STIRAP scheme would require at least three lasers with individual control over each laser's detuning and intensity. In contrast, FFM is simpler to implement and does not require additional optomechanical components beyond what is needed for a monochromatic Rydberg drive.

\clearpage
%%%%%%%%%%%%%%%%%%%%%%%%%%%%%%%%%%%%%%%%%%%%%%%%%%%%%%%%%%%%%%%%%%%%%%%%%%%%%%%%%%%%%%
%\bibliographystyle{sn-nature}
%\bibliography{references}

%%%%%%%%%%%%%%%%%%%%%%%%%%%%%%%%%%%%%%%%%%%%%%%%%%%%%%%%%%%%%%%%%%%%%%%%%%%%%%%%%%%%%%
\def\bibcommenthead{}
\def\bibsection{\section*{Supplementary \refname}}

% \bibliographystyle{naturemag}

%%%%%%%%%%%%%%%%%%%%%%%%%%%%%%%%%%%%%%%%%%%%%%%%%%%%%%%%%%%%%%%%%%%%%%%%

\end{document}